 \documentclass[traditabstract]{aa} % for the abstract without structuration 
\usepackage{graphicx}
\usepackage{natbib}
\usepackage{subfigure}
\usepackage{longtable,lscape}
\usepackage{txfonts}
\begin{document}
\title{PACS-\emph{Herschel}\thanks{\emph{Herschel} is an ESA space observatory with science instruments provided by European-led Principal Investigator consortia and with important participation from NASA.} FIR detections of Lyman-alpha emitters at 2.0$\lesssim$z$\lesssim$3.5}

%     \subtitle{HERSCHEL FIR counterparts of Ly$\alpha$ emitters}

\titlerunning{Stellar populations and \emph{Herschel} FIR-counterparts of Ly$\alpha$ emitters at 2 $\lesssim$ z $\lesssim$ 3.5 in GOODS-South}

\author{
I. Oteo\inst{1,2}
\and
A. Bongiovanni\inst{1,2,3}
\and
A.~M. P\'erez Garc\'{\i}a\inst{1,2,3}
\and
J. Cepa\inst{2,1}
\and
A. Ederoclite\inst{4}
\and
M. S\'anchez-Portal\inst{5,3}
\and
I. Pintos-Castro\inst{1,2}
\and
R. P\'erez-Mart\'inez\inst{6}
\and
B. Altieri\inst{5}
\and
P. Andreani\inst{7}
\and
H. Aussel\inst{8}
\and
S. Berta\inst{9}
\and
A. Cimatti\inst{10}
\and
E. Daddi\inst{8}
\and
D. Elbaz\inst{8}
\and
N. F\"orster Schreiber\inst{9}
\and
R. Genzel\inst{9}
\and
D. Lutz\inst{9}
\and
B. Magnelli\inst{9}
\and
R. Maiolino\inst{11}
\and
A. Poglitsch\inst{9}
\and
P. Popesso\inst{9}
\and
F. Pozzi\inst{12}
\and
E. Sturm\inst{9}
\and
L. Tacconi\inst{9}
\and
I. Valtchanov\inst{5}
}
\offprints{Iv\'an Oteo, \email{ioteo@iac.es}}

\institute{Instituto de Astrof{\'i}sica de Canarias (IAC), E-38200 La Laguna, Tenerife, Spain %1
\and Departamento de Astrof{\'i}sica, Universidad de La Laguna (ULL), E-38205 La Laguna, Tenerife, Spain%2
\and Asociaci\' on ASPID. Apartado de Correos 412, La Laguna, Tenerife, Spain%3
\and Centro de Estudios de F\'isica del Cosmos de Arag\' on, Plaza San Juan 1, Planta 2, Teruel, 44001, Spain%4
\and Herschel Science Centre (ESAC). Villafranca del Castillo, Spain%5
\and XMM/Newton Science Operations Centre (ESAC). Villafranca del Castillo. Spain%6
\and ESO, Karl-Schwarzchild-Str. 2, D-85748 Garching, Germany%7
\and Commissariat \`a l'\'Energie Atomique (CEA-SAp) Saclay, France%8
\and Max-Planck-Institut f\"{u}r Extraterrestrische Physik (MPE), Postfach 1312, 85741 Garching, German%9
\and Dipartimento di Astronomia, Universit\`a di Bologna, Via Ranzani 1, 40127 Bologna, Italy%10
\and INAF - Osservatorio Astronomico di Roma, via di Frascati 33, 00040 Monte Porzio Catone, Italy%11
\and INAF - Osservatorio Astronomico di Bologna, via Ranzani 1, I-40127 Bologna, Italy%12
}

\date{Received ...; accepted... }

% \abstract{}{}{}{}{} 
% 5 {} token are mandatory

 \abstract
   {In this work we analyze the physical properties of a sample of 56 spectroscopically selected star-forming (SF) Ly$\alpha$ emitting galaxies at 2.0$\lesssim$z$\lesssim$3.5 using both a spectral energy distribution (SED) fitting procedure from rest-frame UV to mid-IR and direct 160$\mu$m observations taken with the Photodetector Array Camera \& Spectrometer (PACS) instrument onboard \emph{Herschel Space Observatory}. We define LAEs as those Ly$\alpha$ emitting galaxies whose rest-frame Ly$\alpha$ equivalent widths (Ly$\alpha$ EW$_{rest-frame}$) are above 20\AA, the typical threshold in narrow-band searches. Ly$\alpha$ emitting galaxies with Ly$\alpha$ EW$_{rest-frame}$ are called non-LAEs. As a result of an individual SED fitting for each object, we find that the studied sample of LAEs contains galaxies with ages mostly below 100Myr and a wide variety of dust attenuations, SFRs, and stellar masses. The heterogeneity in the physical properties is also seen in the morphology, ranging from bulge-like galaxies to highly clumpy systems. In this way, we find that LAEs at 2.0$\lesssim$z$\lesssim$3.5 are very diverse, and do not have a bimodal nature, as suggested in previous works. Furthermore, the main difference between LAEs and non-LAEs is their dust attenuation, because LAEs are not as dusty as non-LAEs. On the FIR side, four galaxies of the sample (two LAEs and two non-LAEs) have PACS-FIR counterparts. Their total IR luminosity place all of them in the ULIRG regime and are all dusty objects, with A$_{1200}$$\gtrsim$4mag. This is an indication from direct FIR measurements that dust and Ly$\alpha$ emission are not mutually exclusive. This population of red and dusty LAEs is not seen at z$\sim$0.3, suggesting an evolution with redshift of the IR nature of galaxies selected via their Ly$\alpha$ emission.
   }


   \maketitle
%
%________________________________________________________________

\section{Introduction}\label{intro}

Over the past years, a large number of galaxies have been found at different redshifts, from the local universe up to z$\sim$7 and  even beyond. One of the most successful selection methods for searching for high-redshift galaxies is the narrow-band technique. It employs a combination of narrow and broad band filters to isolate the Ly$\alpha$ emission in the spectrum of a galaxy and to constrain its nearby continuum, respectively. This method segregates galaxies with a Ly$\alpha$ emission whose rest-frame Ly$\alpha$ equivalent width (Ly$\alpha$ EW$_{rest-frame}$) is typically above 20\AA. They are called Ly$\alpha$ emitters (LAEs). 
   
Many efforts have been aimed at looking for LAEs in a wide range of redshifts \citep{Deharveng2008,Cowie2010,Cowie2011, Bongiovanni2010,Guaita2010, Cowiehu1998,Gronwall2007,Gawiser2006,Ouchi2008, Rhoads2000,Shioya2009,Murayama2007,Ouchi2010}. Physical properties of LAEs have been mostly analyzed by fitting their observed spectral energy distribution (SED) with \citet[hereafter BC03]{Bruzual2003} templates. This allows the determination of their dust attenuation, star-formation rate (SFR), age or stellar mass \citep{Finkelstein2008,Finkelstein2009,Finkelstein2009_45, Nilsson2007,Nilsson2009, Nilsson2011,Cowie2011,Ono2010,Lai2008, Pirzkal2007, Gawiser2006, Gawiser2007}. 

BC03 templates do not take the dust emission in the FIR into account and, even if they did, the lack of FIR information for LAEs, mainly at z$\gtrsim$2, would produce important physical parameters, such as dust attenuation and SFR, suffering from large uncertainties. In \cite{Oteo2011_letter,Oteo2011} we look for mid-IR/FIR counterparts of a sample of LAEs at z$\sim$0.3 by using PACS-100$\mu$m, PACS-160$\mu$m, and MIPS-24$\mu$m data, finding that a high percentage of them ($\sim$75\%) are detected at those wavelengths. These detections enable us to determine their IR nature, dust attenuation, and SFR without the intrinsic uncertainties of SED fitting. As a result, we find that LAEs at z$\sim$0.3 are among the least dusty galaxies at that redshift and the majority have total IR luminosities below 10$^{11}$L$_{\odot}$.

%THIS WORK

The 2$\lesssim$z$\lesssim$3.5 range (1.7 to 3.2 Gyr after the Big Bang) is where the SFR density of the Universe is at a maximum \citep{Hoog1998, Hopkins2004, Hopkins2006,Perez2005,Lefloch2005}, and, with respect to LAEs, it has only started to be studied recently. \cite{Nilsson2009} demonstrate that there is a significant evolution in the physical properties of LAEs between z$\sim$3.0 and z$\sim$2.3, with a spread in their SEDs which is greater at z$\sim$2.3 than at z$\sim$3.0. This indicates that LAEs at z$\sim$2.3 are more massive, older, and/or dustier than those at higher redshifts. \cite{Nilsson2011} find that both the dust attenuation and stellar mass of LAEs at z$\sim$2.3 are high, mainly spanning A$_{V}$= 0.0-2.5 mag and log(M$_{*}$/M$_{\odot}$)= 8.5-11.0, respectively. They also find that physical properties of LAEs at that redshift are very diverse. \cite{Guaita2011} obtain robust determinations of mass and dust attenuation, log(M$_{*}$/M$_{\odot}$) = 8.6[8.4-9.1] and E(B-V) = 0.22[0.00-0.31], for LAEs at z$\sim$2.1. Furthermore, comparing with a sample of LAEs at z$\sim$3.1, they find that LAEs at z$\sim$2.1 tend to be dustier and show higher instantaneous SFR than those at z$\sim$3.1, and the properties are also diverse. \cite{Bond2011_evolution} claim between z$\sim$3.1 and $\sim$2.1 for an evolution of the morphological properties of LAEs, with the median half-light radii rising with decreasing redshift. They also report that LAEs at z$\sim$2.1 are bigger for galaxies with higher stellar mass, star formation rate, and dust obscuration. Therefore, 2.0$\lesssim$z$\lesssim$3.0 is an interesting redshift range for several reasons: 1) large samples of LAEs are being collected, 2) in this redshift range, LAEs tend to have bright observable fluxes, which allows a multiwavelength coverage with a higher signal-to-noise ratio than at higher redshifts, 3) it is a redshift range where the physical properties of LAEs seem to be changing significantly and become very diverse, while this has not been clearly reported at higher redshifts.

In this work, we focus on a sample of spectroscopically selected Ly$\alpha$ emitting galaxies at 2$\lesssim$z$\lesssim$3.5. The spectroscopic selection segregates galaxies with Ly$\alpha$ EW$_{rest-frame}$ either above or below 20\AA, the typical threshold in narrow-band searches. To place our contribution in a common base with narrow-band selected galaxies, we distinguish the following throughout the work: \emph{LAEs}: Ly$\alpha$ emitting galaxies with Ly$\alpha$ EW$_{rest-frame}$$>$20\,\AA; \emph{non-LAEs}: Ly$\alpha$ emitting galaxies with Ly$\alpha$ EW$_{rest-frame}$$<$20\,\AA. The main objective of this work is to give a first glimpse at the FIR properties of LAEs at 2.0$\lesssim$z$\lesssim$3.5 by using deep FIR data coming from \emph{Herschel}-PACS observations \citep{Poglitsch2010,Pilbratt2010}. Furthermore, with the aim of better understanding this population in that redshift range, we also study their physical properties by employing a SED fitting procedure with BC03 templates. This allows the comparison between SED fitting and IR-based results.

%STRUCTURE

This paper is divided in two main parts. First, in Sections \ref{data}, \ref{fitting_method}, \ref{stellar_populations}, and \ref{morfo} we describe the optical to mid-IR data used in this work, perform SED fittings with BC03 templates, and carry out a morphological analysis. Then, in Section \ref{FIR_counterparts}, we focus on the FIR side of their SED, studying total IR luminosities, IR nature, total SFR, and dust attenuation for the FIR-detected galaxies. Section \ref{Conclusions} gives the conclusions of this work.

%COSMOLOGY ASSUMED

Throughout this paper we assume a flat universe with $(\Omega_m, \Omega_\Lambda, h_0)=(0.3, 0.7, 0.7)$. All magnitudes are listed in the AB system \citep{Oke1983}.

   \begin{figure}
   \centering
   \includegraphics[width=0.46\textwidth]{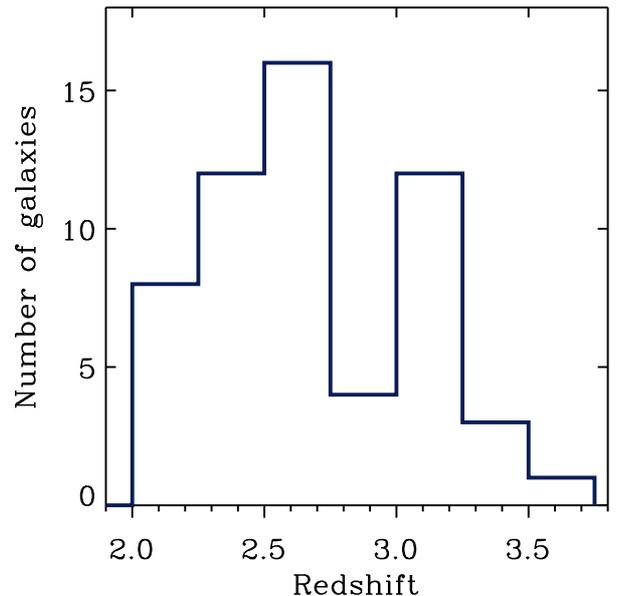}\\
      \caption{Redshift distribution of the sources in the final sample. We study objects at 2 $\lesssim$ z $\lesssim$ 3.5. 
              }
         \label{redshift}
   \end{figure}

%%%%%%%%%%%%%%%%%%%%%%%%%%%%%%%%%%%%%%%%%%%%%%%%%
% OPTICAL DATA
%%%%%%%%%%%%%%%%%%%%%%%%%%%%%%%%%%%%%%%%%%%%%%%%%

\section{Optical data and object selection}\label{data}

The sample used in this work was selected from the final data release of the VIMOS spectroscopic campaign in the GOODS-South field \citep{Popesso2009,Balestra2010}. VIMOS spectra have a quality flag assigned as follows: A (secure classification), B (likely classification), and C (tentative classification). From all the spectra in the survey, we only selected those galaxies that exhibit a Ly$\alpha$ emission in their spectrum and which are flagged as A or B. This yields a sample of 144 objects. The wavelength coverage of the spectra implies that the LAEs are distributed within 2.0$\lesssim$z$\lesssim$3.5. Throughout this work, the objects are named as in \cite{Popesso2009}.

      \begin{figure}
   \centering
   \includegraphics[width=0.46\textwidth]{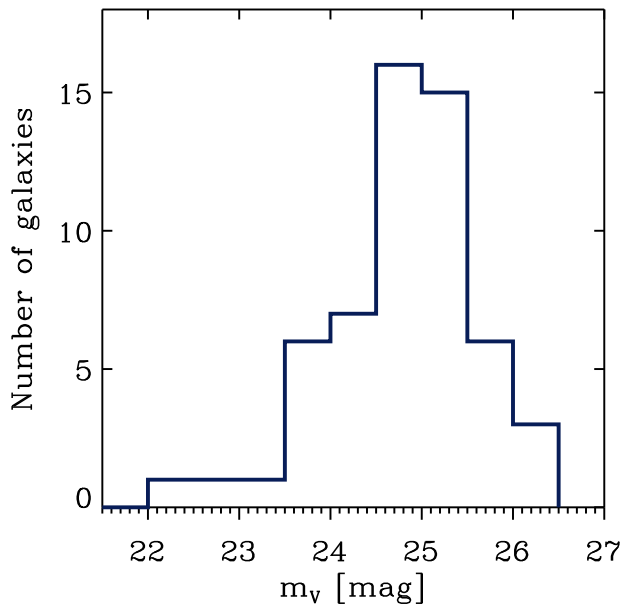}
      \caption{Histogram of the magnitudes in the V band of the final sample of 56 LAEs. The values appearing here are similar to those in the UV-bright sample of \cite{Guaita2011}.
              }
         \label{magV}
   \end{figure}

With the aim of carrying out SED fittings for the studied galaxies, we used the MUSIC multiwavelength photometric catalog (from 0.3 to 24.0 $\mu$m) of a large and deep area in the GOODS-South Field. Its first version was made by \cite{Grazian2006}, and a subsequent update was elaborated by \cite{Santini2009}. The former uses F435W, F606W, F775W, and F850LP ACS images, JHKs ISAAC data, mid-IR data provided by IRAC instrument (3.6, 4.5, 5.8, and 8.0 $\mu$m), and publicly available U-band data from WFI and VIMOS. The z-band of ACS GOODS frames and Ks of VLT images were used to select objects in the field, yielding a unique, autoconsistent catalog. The details of the catalogs (object detections, PSF-matchings, limiting magnitudes, etc.) can be found in \cite{Grazian2006}. The second version includes objects selected from the IRAC-4.5 $\mu$m image and, therefore, sources detected in that wavelength but very faint or undetected in $K_s$ band. MIPS-24$\mu$m photometry was also included, although it will not be considered in the SED fittings (see Section \ref{fitting_method}). Owing to these improvements, we adopted the second version of the photometric catalog. 

To build our final sample of Ly$\alpha$ emitting galaxies, we matched the coordinates of 144 spectra with the MUSIC catalog, resulting in a total of 70 objects. The loss of 50\% of the sources is explained by the area covered by the VIMOS spectroscopic observations being larger than the GOODS-MUSIC footprint. From that common sample, we visually inspected each spectrum, ruling out objects with no clear Ly$\alpha$ emission or with AGN ionization lines in their spectra. We finally ended with a clean sample of 56 Ly$\alpha$ emitting galaxies whose optical spectra suggest that they have a star-forming (SF) nature. Figure \ref{redshift} shows the redshift distribution of the objects in the final sample. It can be seen that most of them are at 2.0$\lesssim$z$\lesssim$2.7, although we also have a significant number of sources at z$\gtrsim$3. Figure \ref{magV} shows the distribution of the observed V-band magnitude of the studied galaxies. Most of them are brighter than 25.5 mag in that band, which is comparable to the brightness of the UV-bright subsample defined in \cite{Guaita2011}. The main advantage of working with a continuum-bright sample is that it is possible to carry out individual SED fitting for each galaxy, avoiding the uncertainties of stacking \citep{Nilsson2011}.

%%%%%%%%%%%%%%%%%%%%%%%%%%%%%%%%%%%%%%%%%%%%%%%%%
% 	SED FITTING PROCEDURE
%%%%%%%%%%%%%%%%%%%%%%%%%%%%%%%%%%%%%%%%%%%%%%%%%

\section{SED fitting method}\label{fitting_method}

We performed SED fittings with BC03 templates for the sample of 56 galaxies by using the Zurich Extragalactic Bayesian Redshift Analyzer \citep[ZEBRA,][]{Feldmann2006}. In its maximum-likelihood mode, ZEBRA employs a $\chi^2$ minimization algorithm over the templates to find the one that fits the observed SED of each object best. In this process, we excluded the MIPS-24$\mu$m measurements, since the fluxes in this band have a significant contribution of warm dust and PAH molecules emission, which are not taken into account in the BC03 templates. We employed \verb+GALAXEV+, which is provided by BC03, to build a large sample of templates. In this process we adopted a \cite{Salpeter1955} initial mass function, distributing stars from 0.1 to 100M$_\odot$. This IMF has mostly been used in previous works in the studied redshift range \citep{Guaita2011,Nilsson2011,Lai2008,Ono2010,Gawiser2006,Gawiser2007}. Since metallicities do not significantly modify the shape of the SEDs of galaxies, we adopted Z=0.4Z$_{\odot}$. Different authors employ different values: \cite{Lai2008} and \cite{Guaita2011} consider the solar metallicity, whereas \cite{Ono2010} utilize a value of 0.2Z$_{\odot}$. To check the validity of our choice, we ran ZEBRA with BC03 templates associated to different metallicities and we find similar results for the other parameters, within the uncertainties. This, at the same time, is an indication that metallicity is difficult to determine accurately with SED fitting. For the SFR, we use models with temporally constant values, as in many previous works \citep{Gawiser2006,Lai2008,Ono2010,Guaita2011,Kornei2010}. In this case, the SFR can be obtained from the rest-frame UV fluxes once the templates are normalized to the observed photometry and employing the \cite{Kennicutt1998} calibration between SFR and the UV luminosity. We also tried to use exponentially decreasing SFR, but the differences between the results for age and dust attenuation were not significant. \cite{Ono2010} also compare the results obtained with constant and varying SFR and find similar results in both cases, which indicates that the history of the star formation in a galaxy is hard to constrain with SED-fitting methods. Furthermore, adopting an exponentially varying SFR would add a new degree of freedom in the process, the star-formation time scale, introducing more uncertainties in the determination of the other parameters.

The ages considered here span from 1 Myr to 3.4 Gyr, the age of the universe at the minimum redshift of the sample, z$\sim$2.0. Dust attenuation is included via the \cite{Calzetti2000} law with values of the color excess in the stellar continuum, E(B-V)$_s$, ranging from 0 to 0.7 in steps of 0.05. We also include intergalactic medium absorption adopting the prescription of \cite{Madau1995}. Stellar masses are obtained from SFR and age, according to the assumed temporal variation of the SFR. Regarding the uncertainties in the fitted values, ZEBRA provides a parameter related to the probability that one template truly represents the observed photometry of a given source. In the SED fitting of each galaxy, we select all the templates whose associated probability is above 68\% and consider the uncertainty of each parameter as the average of the values associated to that parameter on the selected templates with more than 60\% of probability.

%%%%%%%%%%%%%%%%%%%%%%%%%%%%%%%%%%%%%%%%%%%%%%%%%
% STELLAR POPULATIONS
%%%%%%%%%%%%%%%%%%%%%%%%%%%%%%%%%%%%%%%%%%%%%%%%%

\section{Stellar populations}\label{stellar_populations}

\begin{figure*}[!th]
   \centering
   \includegraphics[width=0.8\textwidth]{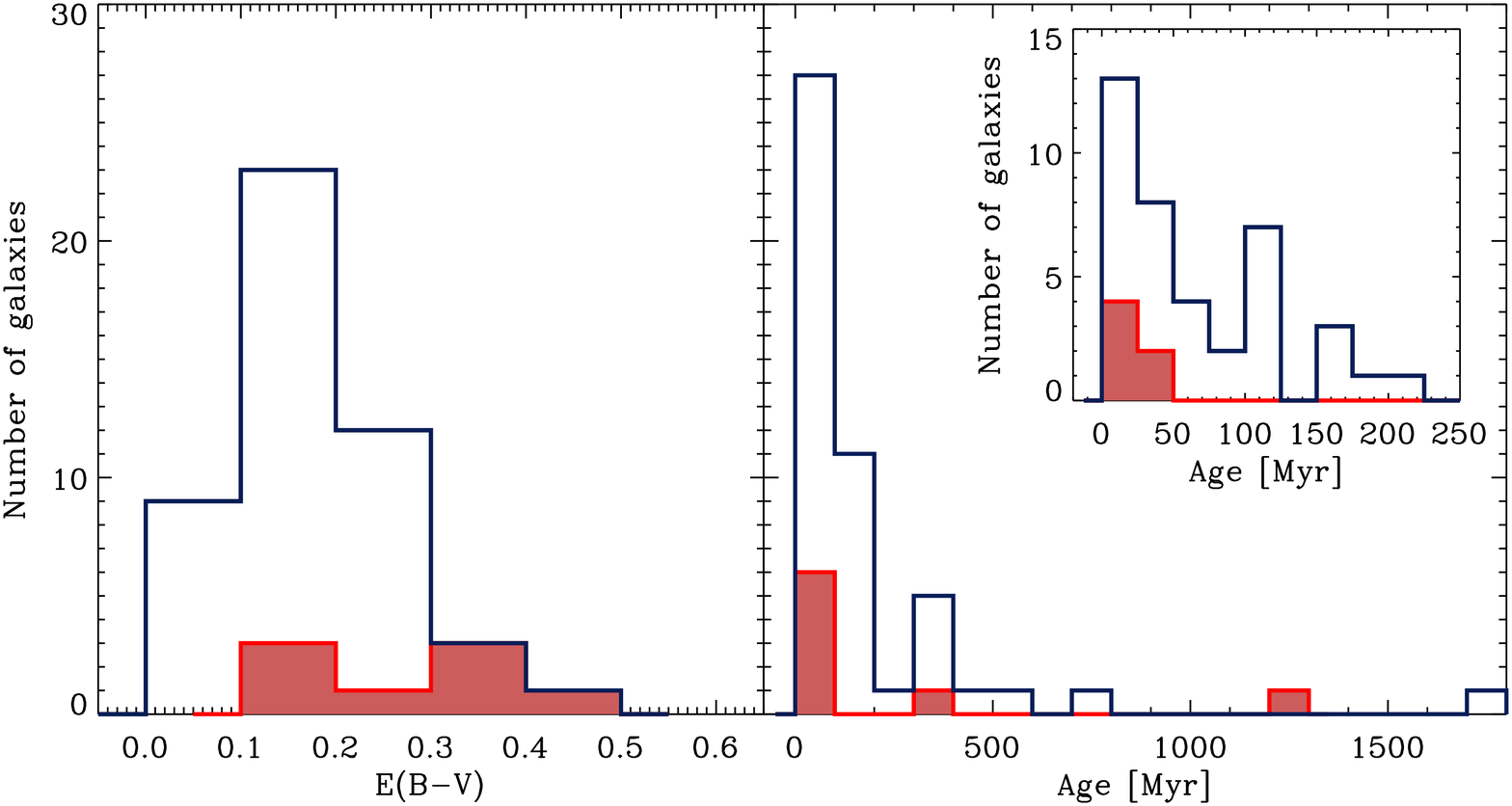}
            \includegraphics[width=0.8\textwidth]{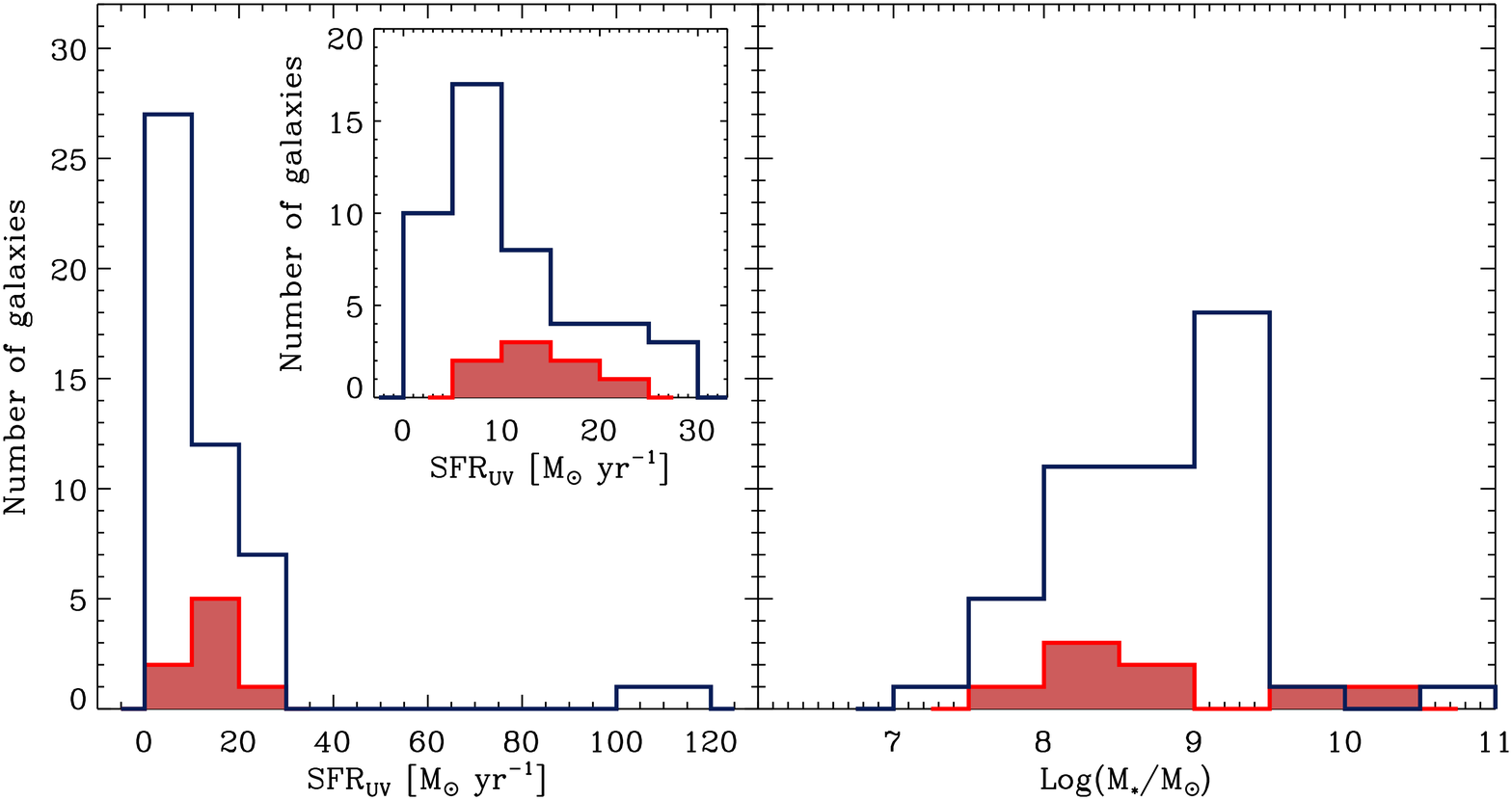}
      \caption{Distribution of the physical properties for the sample of 56 LAEs. Blue and red-shaded histograms represent LAEs and non-LAEs, respectively. Inset plots are more detailed representations of the zones of the histograms where most objects are located.
              }
         \label{edad_polvo}
   \end{figure*}

Figure \ref{edad_polvo} and Table \ref{tabla_LAES_properties} show the results of the SED fitting for the studied galaxies. The dust attenuation of the studied LAEs, parametrized by the color excess in the stellar continuum E(B-V)$_s$, ranges from 0 to 0.3, with a median value of 0.15. The uncertainties in dust attenuation have a lower limit equal to the sampling of this parameter in the templates, that is, $\Delta$E(B-V)$_{s}$$\sim$0.05, and are also affected by the well-known degeneracy between age, dust attenuation, and metallicity of the SED-fitting methods. Despite the uncertainties, the values obtained here are distributed within approximately the same interval as in \cite{Guaita2011} for the UV-bright sample, although their median value, E(B-V)$\sim$0.32, is higher than that found in the present work. 

Regarding age, the studied LAEs are young galaxies, most of them having ages below 100 Myr. However, there are some LAEs, $\sim$20\%, whose SEDs are compatible with older populations, above 300 Myr. Although the threshold age in the generation of the templates was set to the age of the universe at z$\sim$2, there is no violation of the age of the universe for galaxies at higher redshifts. The uncertainties in age are typically below the 20\%, which hints towards robust age estimations.

Intense rest-frame optical emission lines, such [OII], [OIII] or H$\alpha$, could have a strong effect on the observed broadband photometry of a galaxy. At z$\sim$0.3, \cite{Cowie2011} employed their rest-frame optical spectroscopic observations of their LAEs to analyze the influence of rest-frame optical emission lines on the derived parameters in a SED fitting procedure. As a result, it was found that not subtracting those lines overestimates of the ages of the LAEs in their sample. In our work, given that we do not have spectral information about the [OII], [OIII], or H$\alpha$ emission for the studied galaxies, we can not correct for this effect, and, so the ages obtained here should be considered as an upper limit of the real values.

Since the studied LAEs have a bright UV continuum, their dust-uncorrected SFRs, SFR$_{UV,uncorrected}$, are high, ranging from about 2 to 100 M$_\odot$yr$^{-1}$, with a median value of SFR$_{UV,uncorrected}$$\sim$10 M$_{\odot}$yr$^{-1}$. These values are slightly higher than those reported at lower redshifts. In \cite{Oteo2011} we studied the SFR$_{UV,uncorrected}$ for LAEs at z$\sim$0.3, finding that most of them have SFR$_{UV,uncorrected}$ below 5 M$_{\odot}$yr$^{-1}$. In the present work we do not see such low values because of the UV-brightness of the studied LAEs. However, the high values found at 2.0$\lesssim$z$\lesssim$3.5 are not seen at z$\sim$0.3. This could suggest that the upper limit of the SFR$_{UV,uncorrected}$ in LAEs is increasing from 0.3 to z$\gtrsim$2, or similarly, that there is an evolution in the upper limit of the UV luminosity, L$_{UV}$. The surveyed comoving volume in \cite{Oteo2011} and in the present work are similar given that the GALEX observations used in the former cover a much larger area of the sky than those used in the present work because the larger surveyed area at z$\sim$0.3 makes up for the difference in redshifts. \cite{Cowie2011} report a strong evolution in the Ly$\alpha$ luminosity function of LAEs between z$\sim$0.3 and $\sim$1.0, which could be related to the evolution in the upper limit of L$_{UV}$ reported above. This suggests that there is a noticeable change in the UV properties of galaxies selected via their Ly$\alpha$ emission from z$\sim$0.3 to z$\gtrsim$2.0. 

      \begin{figure*}[!t]
   \centering
   \includegraphics[width=0.8\textwidth]{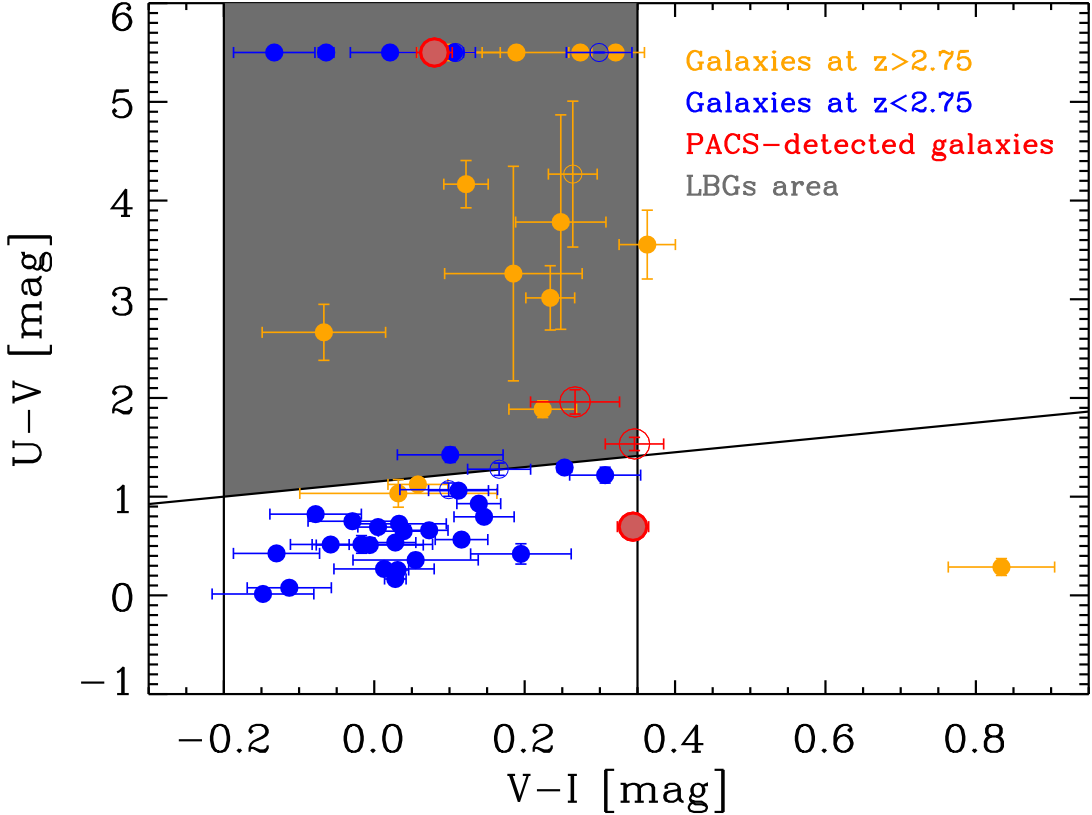}
      \caption{Color-color diagram containing the locus of LBGs according to \cite{Pentericci2010}. Filled and open dots represent LAEs and non-LAEs, respectively. Orange and blue dots represent galaxies at z$>$2.75 and z$<$2.75, respectively. Red dots are PACS-detected galaxies and shaded zone is the area where LBGs are located. For clarity, the galaxies with a U-V color greater than 5.5 are assigned U-V=5.5.
              }
         \label{LBGs}
   \end{figure*}

The stellar mass of the studied LAEs spans within 7.0$\lesssim$Log(M$_*$/M$_\odot$)$\lesssim$9.5. More than 70\% of the studied Ly$\alpha$ emitting galaxies are detected in IRAC-3.6$\mu$m under a similar limiting luminosity to the one in \cite{Lai2008}. In this way, we are working with a sample of IRAC-bright sources, and they should be among the most massive LAEs in the studied redshift range. The range of stellar masses obtained here is compatible with the results of \cite{Guaita2011} for the UV-bright sample and with those at z$\sim$3.0 of \cite{Lai2008} and \cite{Gawiser2006,Gawiser2007} for both IRAC-detected and IRAC-undetected stacked samples. 

After inspecting the individual SED fittings we conclude that there is a wide range of shapes, from almost-dust free cases, with a negative UV continuum slope, to highly attenuated LAEs, with a positive UV continuum slope. The rest-frame optical-to-UV colors also show a significant variation from blue to red objects. This way, analyzing individual SEDs for each object, we find LAEs with very different properties, rather than the double nature suggested in previous works, which could be the result of the stacking method being employed \citep{Lai2008,Gawiser2006,Gawiser2007}. Actually, \cite{Nilsson2011} studied the effects of stacking in their sample of LAEs at z$\sim$2.3, finding that, while stellar masses are robust to stacking, ages and dust attenuation tend to be incorrectly determined. Therefore, it is clear that to analyze the physical properties of LAEs at high redshift, individual fits to the observed photometry are needed, although it is challenging for those LAEs at z$\gtrsim$3 owing to photometric limitations. Actually, only a few LAEs of the samples of \cite{Ono2010} and \cite{Lai2008} are detected in the Ks band, which, at their redshifts, is quite important for sampling the Balmer break and obtaining accurate values of age and mass. Results based on stacking analysis must be taken with care, most importantly in those cases where the SEDs are known to have very diverse shapes and properties, as shown in this and previous works to happen at 2.0$\lesssim$z$\lesssim$3.5 \citep{Guaita2011,Nilsson2011}.

As mentioned in Section \ref{intro}, the spectroscopic selection used in this work allows isolating sources with Ly$\alpha$ EW$_{rest-frame}$$<$20\,\AA\, (non-LAEs). However, the number of such sources in our sample is low. Non-LAEs are also represented in Figure \ref{edad_polvo}. It can be seen that the age, SFR and mass of these galaxies are in the same range as those for LAEs. The main difference is the dust attenuation: objects with an attenuated Ly$\alpha$ emission are among the dustiest objects in the sample. This points towards the idea that LAEs are among the least dusty galaxies at each redshift, as reported in other works for different redshifts \citep{Cowie2011,Oteo2011_letter,Oteo2011,Pentericci2007,Pentericci2010,Kornei2010}.

Figure \ref{LBGs} represents the locus of Lyman Break Galaxies (LBGs) at z$\sim$3.0 for the filters in the MUSIC survey \citep{Pentericci2010}. It can be seen that most LAEs at z$\gtrsim$2.75 could have been also selected as LBGs. Most LAEs at z$\lesssim$2.75 do not meet the LBG criterion, because for those redshifts, the Lyman break is not located between the U and V bands, but in bluer wavelengths. The subsample of LAEs with colors of LBGs (LAE-LBGs) is the bridge population between these two kinds of galaxies. The principal difference between LAE-LBGs and the total sample of LAEs is the dust attenuation: LAE-LBGs tend to be dustier than other LAEs. In Figure \ref{LBGs}, this tendency is reflected in their V-I color. At our redshifts, this color measures the UV continuum slope, and higher values correspond to greater dust attenuation. LAE-LBGs have higher V-I color (redder UV continuum) than other LAEs of the sample, since they are dustier. \cite{Pentericci2010} find that LBGs with a Ly$\alpha$ emission at our redshift have lower dust attenuation than those without a Ly$\alpha$ line. The combination of both results points toward a sequence of increasing dust attenuation, from the less-dusty non-LBG LAEs, to dustier LAE-LBGs, and finally those LBGs without Ly$\alpha$ emission, so they could be the same population differentiated by dust attenuation.

%%%%%%%%%%%%%%%%%%%%%%%%%%%%%%%%%%%%%%%%%%%%%%%%%
% MORPHOLOGY
%%%%%%%%%%%%%%%%%%%%%%%%%%%%%%%%%%%%%%%%%%%%%%%%%

\section{Morphology}\label{morfo}

By using HST/ACS images in GOODS-South we carried out a morphological analysis of the 56 studied Ly$\alpha$ emitting galaxies. We adopted the classification given in \cite{Elmegreen2009}, and visually classify them into three groups: 37 clump clusters (CC), 8 chain galaxies (CH), and 11 spiral-bulges (SPB). This classification was done by three different people and the results were similar in the three cases. There is a wide variety of sizes and morphologies, from highly compact to highly clumpy objects. This heterogeneity in morphology is compatible with the wide range of physical properties for our LAEs reported in the previous section, indicating that galaxies segregated by their Ly$\alpha$ emission at our redshift do not have specific properties, but instead can be very diverse.

\begin{figure}
   \centering
      \includegraphics[width=0.12\textwidth]{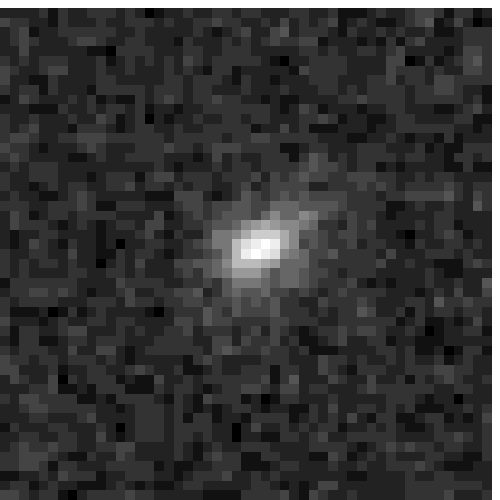}
      \includegraphics[width=0.12\textwidth]{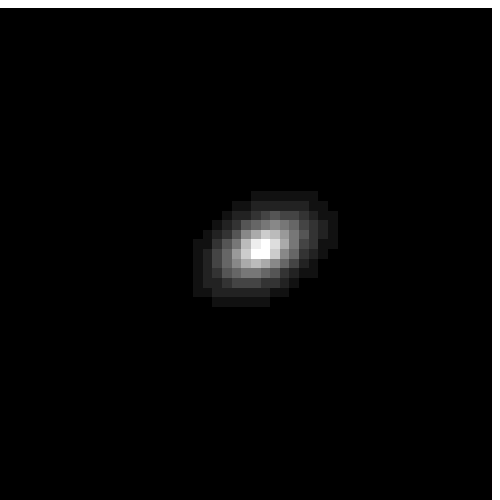}
      \includegraphics[width=0.12\textwidth]{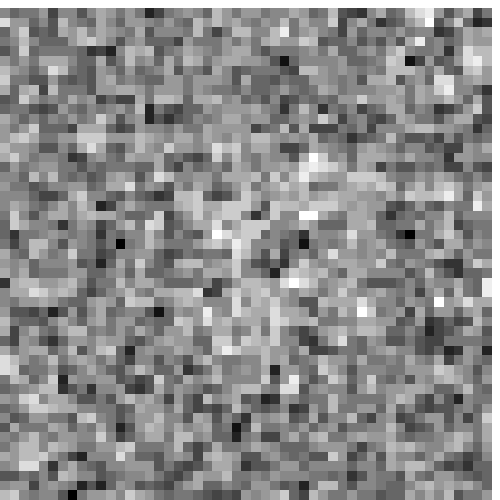}\\

            \includegraphics[width=0.12\textwidth]{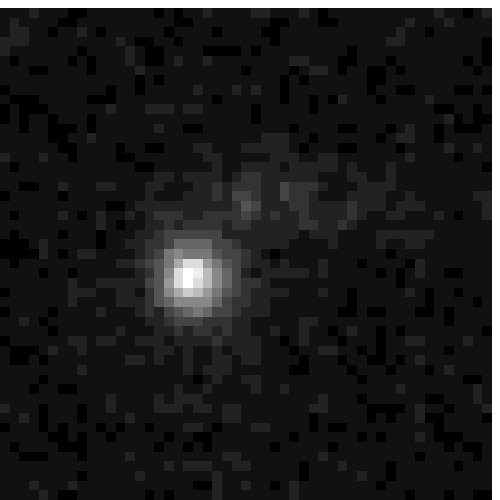}
      \includegraphics[width=0.12\textwidth]{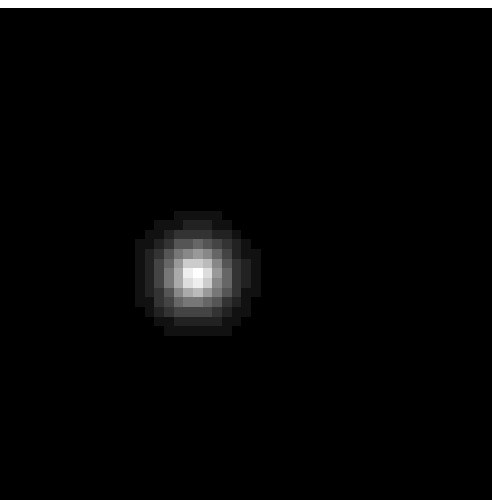}
      \includegraphics[width=0.12\textwidth]{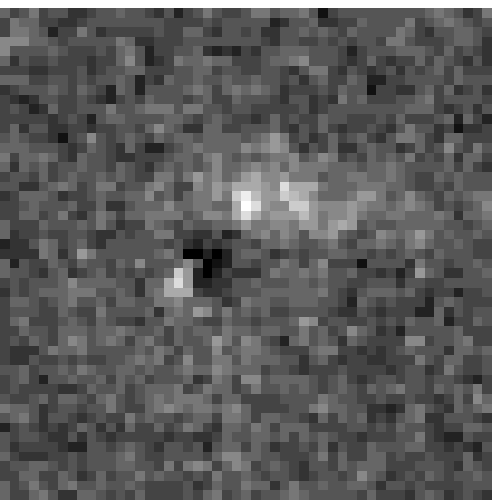}\\

            \includegraphics[width=0.12\textwidth]{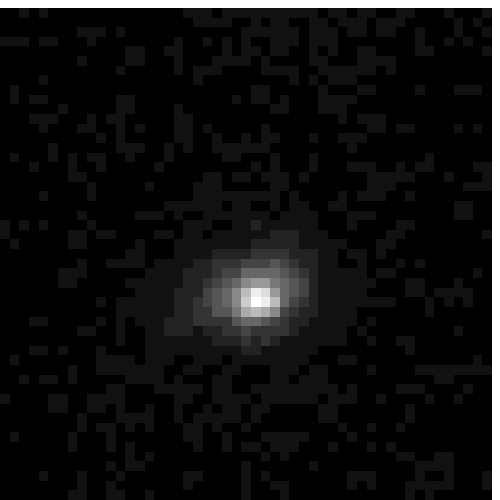}
      \includegraphics[width=0.12\textwidth]{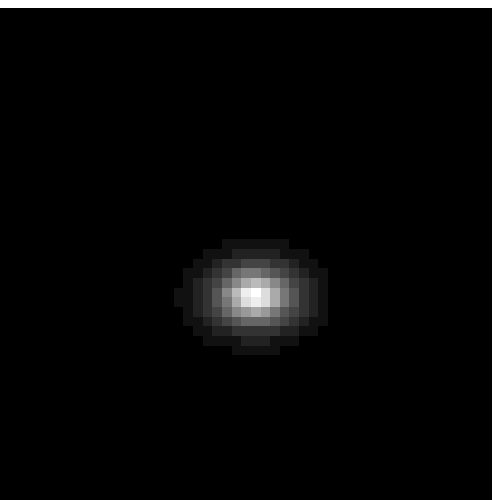}
      \includegraphics[width=0.12\textwidth]{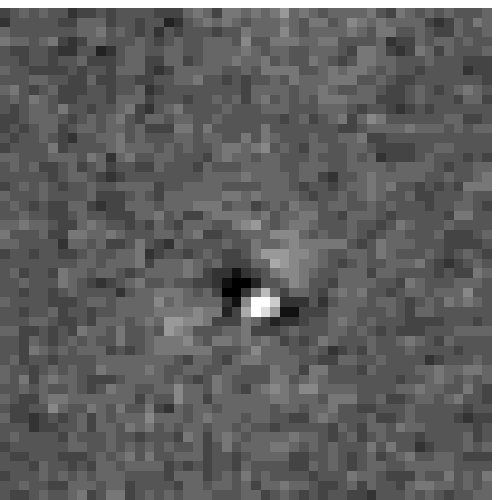}\\ 
      \caption{Examples of GALFIT fittings for three SPB LAEs. \emph{Left}: Images of the LAEs. \emph{Middle}: Sersic fitted profiles. \emph{Right}: Residuals of the fittings.
              }
         \label{GALFIT}
   \end{figure}

We fit the SPB LAEs, which are all at z$<$2.75, with \cite{Sersic1968} profiles by using \verb+GALFIT+ \citep{Peng2010} in the z-band images. At our redshift, this band represents the emission of the rest-frame optical light near the Balmer break. The Sersic profiles can be analytically described by

\begin{equation}
 I(r) = I_b(0) \exp{\left[-b_n \left(r/R_{\textrm{eff}}\right)^{1/n}\right]}
\end{equation}

\noindent where $I_b(0)$ is the central intensity, $R_{\textrm{eff}}$ the effective radius, and $n$ the Sersic index. Figure \ref{GALFIT} shows how GALFIT performs the fittings in three randomly selected SPB LAEs, which are representative of the behavior of the total sample of SPB LAEs. It can be seen that the \cite{Sersic1968} profiles describe the morphology of these objects quite well. To obtain the intrinsic $R_{\textrm{eff}}$ in kpc from the output value given by \verb+GALFIT+ (in pixels), we made use of the assumed cosmology and the ACS pixel scale. As the result, we find that the SPB LAEs have effective radii ranging from 2.6 to 3.1 kpc. The uncertainty in the effective radius, which is provided by \verb+GALFIT+, is less than 5\%. This range is compatible with the tail of the distribution at large effective radii found in \cite{Bond2009} at z$\sim$3.1, in \cite{Bond2011_evolution} at z$\sim$2.1, and in \cite{Oteo2011} at z$\sim0.3$. Therefore, an evolution in the physical sizes of LAEs with redshift is not seen with the present data.

Regarding the Ly$\alpha$ EW$_{rest-frame}$, we do not find any significant difference in the morphology of LAEs and non-LAEs, because of the scarcity of galaxies in the non-LAE sample. At z$\sim$0.3, in \cite{Oteo2011} we find that LAEs and non-LAEs have a clear difference in their morphology, both in shape and size, which indicates that at that redshift Ly$\alpha$ photons are escaping from small and irregular/merging galaxies. However, \cite{Pentericci2010} show that at our redshift, the rest-frame UV morphology does not depend strongly on the presence of the Ly$\alpha$ emission.

%%%%%%%%%%%%%%%%%%%%%%%%%%%%%%%%%%%%%%%%%%%%%%%%%
%           FIR
%%%%%%%%%%%%%%%%%%%%%%%%%%%%%%%%%%%%%%%%%%%%%%%%%

\section{\emph{Herschel} FIR counterparts}\label{FIR_counterparts}

\begin{figure}
   \centering
   \includegraphics[width=0.47\textwidth]{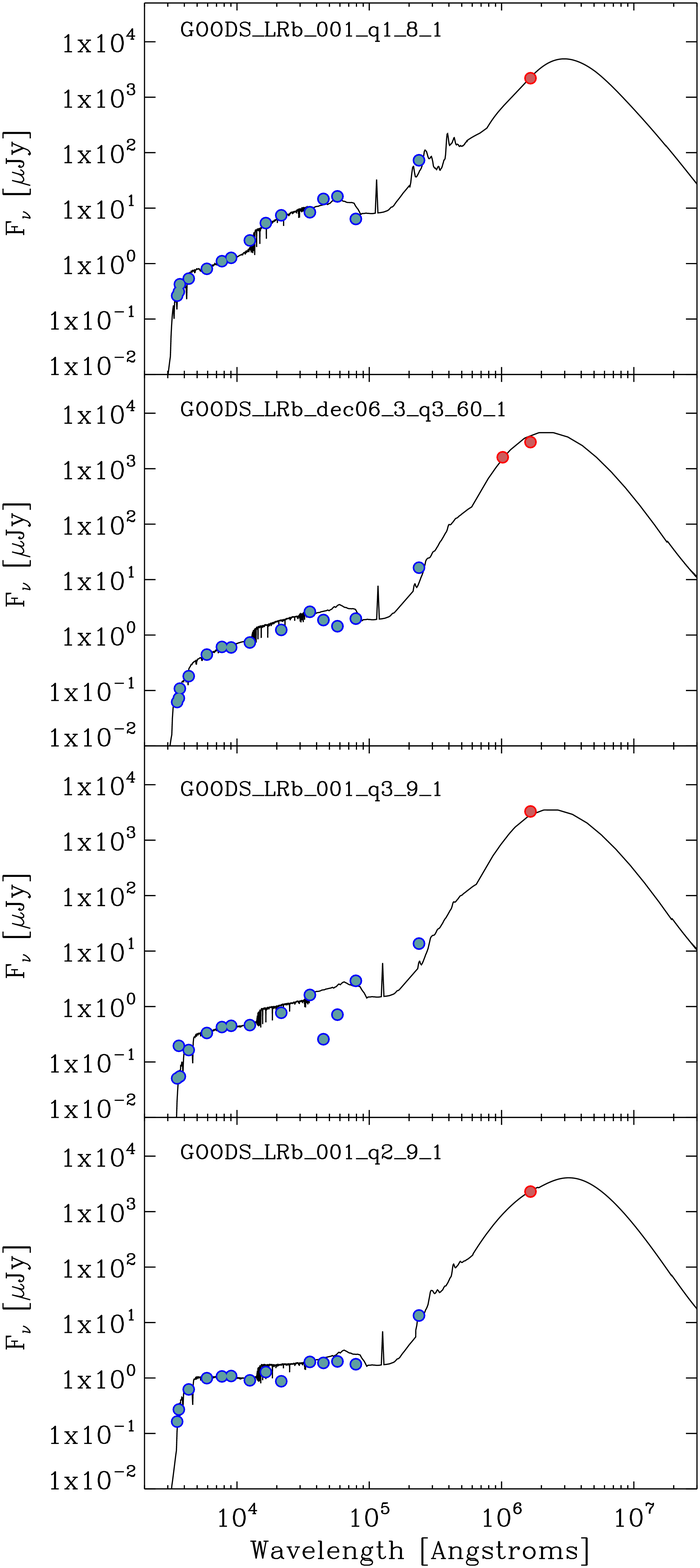}
      \caption{Optical-to-FIR observed SED of the PACS-detected galaxies. Blue dots represent the multi-wavelength photometry from U-band to MIPS-24$\mu$m, and red dots are the PACS-100$\mu$m and/or PACS-160$\mu$m fluxes. Black curve represents the combination of BC03 and CE01 templates that best fit the photometric data of each galaxy.
              }
         \label{FIR}
   \end{figure}

GOODS-South was observed with PACS-70$\mu$m, PACS-100$\mu$m, and PACS-160$\mu$m in the frame of the PACS Evolutionary Probe project (PEP, PI D. Lutz). PEP is the \emph{Herschel} Guaranteed Time Key-Project designed to obtain the best profit from \emph{Herschel} instrumentation to study the FIR galaxy population \citep{Lutz2011}. PACS fluxes used in this work were extracted using MIPS-24$\mu$m position priors with, at least, a 3$\sigma$ significance. Limiting fluxes in PACS-70$\mu$m, PACS-100$\mu$m, and PACS-160$\mu$m are 1.0mJy, 1.1mJy, and 2.0mJy, respectively. We look for possible FIR counterparts of our galaxies within 2'', which is the typical astrometric uncertainty in the position of the sources, finding four detections in at least the PACS-160$\mu$m band. These counterparts are direct evidence of dust emission in high-redshift SF sources. As seen in Figs. \ref{morfo1}, \ref{morfo2} and \ref{morfo3}, GOODS\_LRb\_001\_q2\_9\_1, GOODS\_LRb\_001\_q3\_9\_1, and GOODS\_LRb\_001\_q1\_8\_1 are isolated sources and, therefore, the dust emission is coming entirely from these sources. However, GOODS\_LRb\_dec06\_3\_q3\_60\_1 has a nearby galaxy whose emission in the FIR could contaminate the FIR flux of this galaxy. Two of the PACS-detected galaxies are within the LAE group and the other two have Ly$\alpha$ EW$_{rest-frame}$ compatible with their being non-LAEs. Figure \ref{FIR} represents the rest-frame UV to FIR SED of the four PACS-detected galaxies.

Table \ref{table_L_IR} shows the dust attenuation and SFR$_{UV,uncorrected}$ of the four PACS-detected galaxies, as derived from SED fitting. Their rest-frame UV to mid-IR SEDs are compatible with high values of dust attenuation, and the two LAEs also exhibit greater SFR$_{UV,uncorrected}$ than the median value of the sample. Their large SFR$_{UV,uncorrected}$ indicates that they have high value of UV luminosity, compatible with the greater likelihood that UV-bright galaxies are detected in the FIR \citep{Reddy2010}. GOODS\_LRb\_001\_q2\_9\_1, GOODS\_LRb\_001\_q3\_9\_1, and GOODS\_LRb\_dec06\_3\_q3\_60\_1 are young objects with masses similar to the median population. However, GOODS\_LRb\_001\_q1\_8\_1 is an old and dusty LAE, and it is among the most massive objects of the sample. This is one of the most interesting galaxies in our sample, since its physical properties are opposite the classical idea of an LAE, i.e. a young, less massive, and dust-free galaxy. Three out of the four PACS-detected galaxies would have been selected via the Lyman break technique (see Figure \ref{LBGs}). The other PACS-detected source would have not been segregated as LBG either, since its redshift is not high enough to place the Lyman break between the U and V filters. It can be seen that PACS-detected sources have among the highest V-I colors of the total sample, indicating that they have a high-attenuated UV continuum, compatible with being among the dustiest galaxies in our sample.

%%%%%%%%%%%%%%%%%%%%%%%%%%%%%%%%%%%%%%%%%%%%%%%%%
%           IR  luminosities
%%%%%%%%%%%%%%%%%%%%%%%%%%%%%%%%%%%%%%%%%%%%%%%%%

\subsection{IR luminosities}\label{lumi_IR}

Accurate values of the total 8-1000$\mu$m IR luminosities, L$_{IR}$, can be obtained for the PACS-detected galaxies. Since we do not have full coverage of the dust emission peak, we cannot calculate L$_{IR}$ by direct integration of the FIR SED. Instead, we convert PACS fluxes into L$_{IR}$ by using \cite{Chary2001} (hereafter CE01) templates, which have only one solution for L$_{IR}$ for each FIR flux and redshift. Obtaining L$_{IR}$ by using single FIR band extrapolations has been employed with successful results in previous works in galaxies at similar redshifts \citep{Nordon2010,Elbaz2010,Elbaz2011}. In fact, \cite{Elbaz2010} combine PACS and SPIRE measurements to show that fitting CE01 templates to the PACS-160$\mu$m flux provides reliable estimates of L$_{IR}$ at our redshift. The typical uncertainties of the derived values are related to the accuracy of single band extrapolations rather than to the uncertainties of the PACS fluxes themselves, and are typically less than 0.2 dex \citep{Elbaz2011}. Actually, for galaxies in our redshift range, PACS-160$\mu$m band extrapolations give the most accurate results for single band extrapolations \citep{Elbaz2010}.

%%%%%%%%%%%%%%%%%%%%%%%%%%%%%%%%%%%%%%%%%%%%%%%%%
%           ULIRGs
%%%%%%%%%%%%%%%%%%%%%%%%%%%%%%%%%%%%%%%%%%%%%%%%%

\subsection{LAE-ULIRGs relation}\label{PACS_ULIRGS}

Previous results suggest that some LAEs at our redshift range are also ultraluminous infrared galaxies (ULIRGs) \citep{Nilsson2009_letter,Nilsson2011,Ono2010,Chapman2005}, but direct detections in the FIR around the dust emission peak have not been reported yet. In the present work, according to their L$_{IR}$, the four PACS-detected galaxies have an ULIRG nature. Owing to the depth of the PACS observations used in this work, the limiting L$_{IR}$ is about 10$^{12}$ L$_{\odot}$ at z$\sim$2.2. Therefore, there could be some PACS-undetected LAEs in our sample, the dustiest at the highest redshifts, that fall in the ULIRG class. Actually, given that FIR detections segregate dusty galaxies, it is expected that the dustiest LAEs, E(B-V)$\gtrsim$0.2 of the sample have an ULIRG nature even when PACS-undetected. Thus, more LAEs with an ULIRG nature should be expected. The number of LAEs with an ULIRG nature found, although scarce, is very important. In \cite{Oteo2011_letter,Oteo2011} we studied the FIR properties of a sample of 23 mid-IR/FIR detected LAEs at z$\sim$0.3, finding that most LAEs are in the normal SF galaxy regime, L$_{IR}$$<$10$^{11}$L$_{\odot}$, with only one having L$_{IR}$$\sim$10$^{11.5}$L$_{\odot}$, and none above that value. The discovery of LAEs with an ULIRG nature at z$\sim$2.4-2.8 suggests that the IR properties of LAEs evolve from z$\sim$0.3 to $\sim$2.5 in the sense that there is a population of red and dusty LAEs that is not seen at lower redshifts. Therefore, we find from direct FIR detections that the ULIRG fraction in LAEs decreases from z$\sim$2.4-2.8 to $\sim$0.3, following the trend found for IR-detected galaxies between those redshifts \citep{Lefloch2005}.

This possible evolution of the IR emission, along with the reported evolution in the UV luminosity (Section \ref{stellar_populations}) and in the Ly$\alpha$ luminosity function from z$\sim$1.0 \citep{Cowie2011}, indicates that either the Ly$\alpha$ selection technique does not trace the same kind of objects with redshift, or the properties of Ly$\alpha$ selected galaxies change with redshift.

%%%%%%%%%%%%%%%%%%%%%%%%%%%%%%%%%%%%%%%%%%%%%%%%%
%           Dust attenuation
%%%%%%%%%%%%%%%%%%%%%%%%%%%%%%%%%%%%%%%%%%%%%%%%%

\subsection{Dust attenuation}\label{dust_FIR}

The ratio between IR and UV luminosities is the best way to obtain the dust attenuation in galaxies. Adopting the \cite{Buat2005} calibration, the dust attenuation for our PACS-detected Ly$\alpha$ emitting galaxies can be obtained with the expression

\begin{equation}\label{A_NUV}
A_{NUV} = -0.0495x^3 + 0.4718x^2 + 0.8998x + 0.2269
\end{equation}

\noindent where, $x=\log{\left(L_{IR}/L_{NUV}\right)}$. The conversion from dust attenuation in NUV band into the color excess in the stellar continuum is carried out by using the \cite{Calzetti2000} reddening law. The values obtained are shown in Table \ref{table_L_IR}. PACS-detected LAEs have high values of dust attenuation and, since they are selected via their IR emission, those values represent an upper limit in dust attenuation of LAEs in the studied redshift range. This is a direct indication from FIR measurements that the Ly$\alpha$ emission and the presence of dust are not mutually exclusive. This result could reinforce the scenario where the Ly$\alpha$ emission can be enhanced under a suitable geometry \citep{Neufeld1991}, for which evidence has already been found at higher redshifts \citep{Finkelstein2008}, but not at our redshift or lower \citep{Guaita2011,Finkelstein2011_lowz}. The dust attenuation derived from SED-fitting tends to be lower than obtained from direct IR/UV measurements. The ULIRG nature plays an important role in this underestimation, since for such as sources, the IR emission is more prominent than in other less IR luminous galaxies, and its inclusion in the analysis is essential to obtaining accurate results.

%%%%%%%%%%%%%%%%%%%%%%%%%%%%%%%%%%%%%%%%%%%%%%%%%
%           SFR
%%%%%%%%%%%%%%%%%%%%%%%%%%%%%%%%%%%%%%%%%%%%%%%%%

\subsection{SFR}

The combination of UV and IR luminosities provides the best determination of the SFR in galaxies. Assuming that all the light absorbed in the UV is in turn reradiated in the FIR, the total SFR of a galaxy, SFR$_{total}$, can be obtained as the sum of SFR$_{UV,uncorrected}$ plus a correction term associated to the dust emission in the FIR. Adopting the \cite{Kennicutt1998} calibration, that correction term can be written as

\begin{equation}\label{SFR_FIR}
\textrm{SFR}_{IR}[M_{\odot}\textrm{yr}^{-1}] = 4.5\cdot10^{-44}L_{IR}\,[\textrm{erg}\,\textrm{s}^{-1}]
\end{equation}

\noindent where L$_{IR}$ is defined in the same way than in Section \ref{lumi_IR}. In this way, SFR$_{total}$ can be inferred from

\begin{equation}\label{SFR_total}
\textrm{SFR}_{total} = \textrm{SFR}_{UV, uncorrected} + \textrm{SFR}_{IR}\, .
\end{equation}

The values of SFR$_{total}$ obtained for the PACS-detected galaxies are also shown in Table \ref{table_L_IR}. Such sources have high SFR$_{total}$, all above 200M$_{\odot}$yr$^{-1}$. PACS-deteced sources are the reddest ones of the sample and the fastest star formers. Although the number of such kind of galaxies is low, their SFRs can therefore be considered as the maximum value of SFRs for LAEs in this redshift range. It can be also seen that the contribution of the SFR$_{IR}$ to the SFR$_{total}$ is quite high, above 93\% in all cases, and, therefore, SED fitting with BC03 templates, which do not take the dust emission in the FIR into account, systematically underestimates this quantity. Since we do not have a statistically significant sample of LAEs detected in the FIR, we cannot evaluate a typical factor of underestimation for LAEs in our redshift range. In \cite{Oteo2011} we found that, for LAEs at z$\sim$0.3, the contribution of the FIR emission is more than 60\% in most cases, so that SFR based on SED fitting is underestimating the SFR by a factor greater than 2. According to the IR evolution reported in Section \ref{PACS_ULIRGS}, it could be expected that the FIR contribution can be higher than 60\% for a high percentage of the sample. Actually, LAEs at z$\sim$2.0 with L$_{IR}$$\gtrsim$10$^{10.5}$L$_{\odot}$ are unlikely to be detected with the current FIR surveys, and their IR contribution to the SFR$_{total}$, assuming a SFR$_{UV,uncorrected}$$\sim$10M$_{\odot}$yr$^{-1}$, is about 50\%.

By using the dust attenuation derived from SED fitting, we can correct the SFR$_{UV,uncorrected}$ of the PACS-detected sources to estimate the SFR$_{total}$. Their values are also shown in Table \ref{table_L_IR}. It can be seen that the SFR$_{total}$ obtained in such a way are much lower than the values obtained from the IR emission. This agrees with \cite{Wuyts2011}, who find that SED fitting tend to underestimate the total amount of star formation in IR-bright galaxies at z$\gtrsim$2.5.

\begin{table*}
\caption{Dust attenuation as derived from SED fitting, E(B-V)$_s$ [BC03], and from the IR/UV ratio, E(B-V)$_s$ [IR/UV], dust-uncorrected SFR, SFR$_{UV,uncorr}$, dust-corrected SFR, SFR$_{dust-corr}$, and total SFR, SFR$_{total}$, for the four PACS-detected sources.}             % title of Table
\label{table:1}      % is used to refer this table in the text
\centering                          % used for centering table
\begin{tabular}{l c c c c c}        % centered columns (4 columns)
\hline\hline                 % inserts double horizontal lines
Object & E(B-V)$_s$ [BC03] & E(B-V)$_s$ [IR/UV] & SFR$_{UV,uncorr}$ [M$_{\odot} \textrm{yr}^{-1}$]  & SFR$_{dust-corr}$ [M$_{\odot} \textrm{yr}^{-1}$]   & SFR$_{total}$ [M$_{\odot} \textrm{yr}^{-1}$] \\
\hline
  GOODS\_LRb\_001\_q1\_8\_1      		& 0.20 	& 0.40	& 19.6 & 	74.9	&	242.3		\\
  GOODS\_LRb\_dec06\_3\_q3\_60\_1  	& 0.4 	& 0.48	& 11.9 &	172.7&	311.2		\\
  GOODS\_LRb\_001\_q3\_9\_1      		& 0.30 	& 0.57	& 10.7 &	80.1	&	452.8		\\
  GOODS\_LRb\_001\_q2\_9\_1		& 0.15 	& 0.42	& 25.6 & 	70.3	&	341.5 		\\    
  \hline                                   %inserts single line
\end{tabular}\label{table_L_IR}
\end{table*}

\begin{table}
\caption{Total infrared luminosities for the MIPS-24$\mu$m detected LAEs. PACS-detected LAEs, which are also detected in MIPS-24$\mu$m are not included in this table.}             % title of Table
\label{table:1}      % is used to refer this table in the text
\centering                          % used for centering table
\begin{tabular}{l c c}        % centered columns (4 columns)
\hline\hline                 % inserts double horizontal lines
Object	&	Redshift	& 	Log(L$_{IR}$/L$_{\odot}$) \\
\hline
  GOODS\_LRb\_002\_1\_q3\_8\_1	&	2.296	&	12.34		\\      
  GOODS\_LRb\_002\_1\_q2\_60\_1	&	2.181	&	11.30		\\      
  GOODS\_LRb\_dec06\_2\_q2\_12\_1	&	2.318	&	12.10		\\      
  GOODS\_LRb\_dec06\_3\_q1\_21\_1	&	2.230	&	11.97		\\      
  GOODS\_LRb\_002\_1\_q4\_66\_1	&	2.586	&	12.29		\\      
  GOODS\_LRb\_dec06\_3\_q4\_53\_1	&	2.702	&	11.62		\\      
\hline                                   %inserts single line
\end{tabular}\label{table_MIPS}
\end{table}

\subsection{MIPS-24$\mu$m counterparts of LAEs}

In addition to the analyzed measurements in PACS-160$\mu$m, six other PACS-undetected galaxies of the total sample of 144 have MIPS-24$\mu$m detections in the FIR regime, which were found following the same criterion as used for PACS counterparts. All these galaxies are classified as LAEs. None of them are within the MUSIC footprint, so we cannot analyze the distinctive properties of MIPS-24$\mu$m-detected LAEs regarding the rest-frame UV to mid-IR SED fitting. The MIPS-24$\mu$m detections also indicate the presence of IR-bright, i.e. dusty, LAEs. At 2.0$\lesssim$z$\lesssim$3.5, the MIPS-24$\mu$m band has a significative contribution of PAH molecule emission, which does not allow us to determine the L$_{IR}$ for MIPS-detected LAEs. Actually, \cite{Elbaz2010} study the validity of MIPS-24$\mu$m extrapolations to the L$_{IR}$, finding that they are valid up to z$\sim$1.5 and for objects that are below the ULIRG limit. At z$\gtrsim$1.5, MIPS-24$\mu$m extrapolations tend to overestimate the L$_{IR}$ in a factor of about 2 or 3. Despite this overestimation, we show the L$_{IR}$ luminosities for the MIPS-24$\mu$m detected LAEs in Table \ref{table_MIPS} as inferred from MIPS-24$\mu$m single band extrapolation, in a similar way to Sect. \ref{lumi_IR} for PACS-detected sources. Owing to the PACS catalogs used in Section \ref{FIR_counterparts}, PACS-detected LAEs are also detected in MIPS-24$\mu$m but we do not consider those objects here because they have been more accurately studied in previous sections. It can be seen in Table \ref{table_MIPS} that, although overestimated, MIPS-24$\mu$m detected LAEs have high values of L$_{IR}$, placing them in, at least, the LIRG regime, L$_{IR}$$>$10$^{11}$L$_{\odot}$. Therefore, in total we gather a sample of ten red and dusty LAEs, which are opposite to the classical idea of LAEs as dust-free galaxies. 

\section{Summary and conclusions}\label{Conclusions}

In this work we have analyzed the rest-frame UV to IR SED of a sample of 56 Ly$\alpha$ emitting galaxies at 2.0$\lesssim$z$\lesssim$3.5. According to their Ly$\alpha$ EW$_{rest-frame}$ we distinguish between LAEs and non-LAEs as those Ly$\alpha$ emitting galaxies with Ly$\alpha$ EW$_{rest-frame}$ above and below 20\AA, respectively. This value is the typical threshold in narrow-band searches. Our main conclusions are as follows.

   \begin{enumerate}

\item Individual SED fittings for the studied LAEs indicate that they are mostly young galaxies with ages below 100 Myr, although some of them have SEDs compatible with populations older than 300 Myr. The dust attenuation ranges from low, E(B-V)$\sim$0.0, to high values, E(B-V)$\sim$0.3, with a median of E(B-V) of 0.15. The SFRs, as derived from UV dust-uncorrected measurements, range from about 2 to 100 M$_{\odot}$yr$^{-1}$, mostly having SFR$_{UV,uncorrected}$$\sim$10M$_{\odot}$yr$^{-1}$.

\item We find that LAEs at 2.0$\lesssim$z$\lesssim$3.5 show a wide range of properties, rather than having a double nature as suggested in previous works. This result has been possible to achieve because our LAEs are continuum-bright objects and, therefore, an individual SED fitting for each source could be carried out. 

\item The variety of the physical properties of LAEs is also reflected in their morphology. We find many different structures, from bulge-like galaxies to highly clumpy systems or chain galaxies. Fitting the bulge-like LAEs to Sersic profiles, we found sizes distributed within 2.6 and 3.1 kpc, compatible with those reported in other works at higher, similar and lower redshifts.

\item We find four Ly$\alpha$ emitting galaxies with PACS-FIR counterparts, two LAEs, and two non-LAEs. This indicates that some LAEs have such high dust content that their emission can be directly detected. Their rest-frame UV to mid-IR SEDs are compatible with objects with high dust attenuation. FIR-detected LAEs are objects with high SFR, above 200M$_{\odot}$yr$^{-1}$, and their L$_{IR}$ classify them as ULIRGs. LAEs with an ULIRG nature are not seen at z$\sim$0.3, suggesting that there is an evolution in their IR emission with redshift. This, along with the rapid evolution of the Ly$\alpha$ luminosity function from z$\sim$0.3 to $\sim$1.0 reported in previous works and the evolution in the L$_{UV}$ from z$\sim$0.3 to z$\gtrsim$2.0 found here indicates that either the physical properties in the UV, optical, and IR of objects that are selected via their Ly$\alpha$ emission change with redshift or else the Ly$\alpha$ selection technique would segregate different types of galaxies at different redshifts.
      
\item We obtain the dust attenuation in the PACS-detected LAEs with MUSIC counterparts by combining UV and FIR information, without the uncertainties of the SED-fitting based methods. The results obtained in this way show that they have color excesses  greater than 0.30 in their stellar continuum. This indicates that PACS-detected LAEs are dusty objects, despite having a Ly$\alpha$ line emission in their spectra. This confirms from direct FIR observations that dust and Ly$\alpha$ emission are not mutually exclusive.

   \end{enumerate}

\begin{acknowledgements}
This work was supported by the Spanish Plan Nacional de Astrononom\'ia y Astrof\'isica under grant AYA2008-06311-C02-01. Based on observations made with ESO Telescopes at the La Silla or Paranal Observatories under programme ID 171.A-3045. PACS has been developed by a consortium of institutes led by MPE (Germany) and including UVIE (Austria); KUL, CSL, IMEC (Belgium); CEA, OAMP (France); MPIA (Germany); IFSI, OAP/AOT, OAA/CAISMI, LENS, SISSA (Italy); IAC (Spain). This development has been supported by the funding agencies BMVIT (Austria), ESA-PRODEX (Belgium), CEA/CNES (France), DLR (Germany), ASI (Italy) and CICYT/MICINN (Spain). The Herschel spacecraft was designed, built, tested, and launched under a contract to ESA managed by the Herschel/Planck Project team by an industrial consortium under the overall responsibility of the prime contractor Thales Alenia Space (Cannes), and including Astrium (Friedrichshafen) responsible for the payload module and for system testing at spacecraft level, Thales Alenia Space (Turin) responsible for the service module, and Astrium (Toulouse) responsible for the telescope, with in excess of a hundred subcontractors.
\end{acknowledgements}

\Online
\appendix

\section{Physical properties of the studied Ly$\alpha$ emitting galaxies}\label{tablita}

\begin{table*}
\caption{Physical properties of our sample of 56 galaxies at 2.0$\lesssim$z$\lesssim$3.5 as derived from SED fitting with BC03 templates.}             % title of Table
\label{table:1}      % is used to refer this table in the text
\centering                          % used for centering table
\begin{tabular}{l c c c c c}        % centered columns (4 columns)
\hline\hline                 % inserts double horizontal lines
Name & Redshift & Age [Myr] & E(B-V)[BC03] & SFR$_{UV,uncorrected}$ [M$_{\odot}$yr$^{-1}$] & log(M$_{*}$/M$_{\odot}$) \\
\hline
                GOODS\_LRb\_dec06\_3\_q2\_33\_2 &      2.0214 &    29 $\pm$    34 &     0.25 $\pm$     0.07 &     6.7 &     8.2 \\
                 GOODS\_LRb\_002\_q3\_86\_1 &      2.0526 &    21 $\pm$     6 &     0.30 $\pm$     0.05 &    17.0 &     8.5 \\
                 GOODS\_LRb\_002\_1\_q3\_88\_1 &      2.0600 &    47 $\pm$     8 &     0.15 $\pm$     0.05 &     4.8 &     8.3 \\
                 GOODS\_LRb\_002\_1\_q4\_79\_1 &      2.1449 &    25 $\pm$     6 &     0.15 $\pm$     0.23 &    20.0 &     8.6 \\
                 GOODS\_LRb\_001\_q3\_24\_2 &      2.1659 &    15 $\pm$     6 &     0.20 $\pm$     0.06 &     9.1 &     8.1 \\
                 GOODS\_LRb\_dec06\_1\_q2\_41\_1 &      2.1809 &   541 $\pm$     6 &     0.10 $\pm$     0.05 &     2.3 &     9.1 \\
                 GOODS\_LRb\_dec06\_1\_q4\_15\_2 &      2.2128 &   121 $\pm$    17 &     0.10 $\pm$     0.05 &    20.4 &     9.3 \\
                 GOODS\_LRb\_001\_q2\_7\_1 &      2.2158 &   383 $\pm$    10 &     0.15 $\pm$     0.05 &     2.5 &     8.9 \\
                 GOODS\_LRb\_dec06\_1\_q2\_48\_1 &      2.2991 &    11 $\pm$    32 &     0.20 $\pm$     0.09 &     3.2 &     7.5 \\
                GOODS\_LRb\_002\_1\_q1\_16\_2 &      2.3115 &     5 $\pm$     5 &     0.25 $\pm$     0.19 &    15.9 &     7.9 \\
                GOODS\_LRb\_002\_1\_q4\_29\_1 &      2.3143 &    11 $\pm$    24 &     0.30 $\pm$     0.06 &     9.7 &     8.0 \\
                GOODS\_LRb\_dec06\_3\_q3\_51\_1 &      2.3166 &   305 $\pm$     9 &     0.15 $\pm$     0.05 &     6.2 &     9.2 \\
                GOODS\_LRb\_dec06\_3\_q2\_22\_1 &      2.3170 &    31 $\pm$    21 &     0.15 $\pm$     0.05 &    17.3 &     8.7 \\
                GOODS\_LRb\_dec06\_1\_q1\_11\_3 &      2.3200 &    55 $\pm$    22 &     0.20 $\pm$     0.06 &    27.5 &     9.1 \\
                GOODS\_LRb\_dec06\_1\_q1\_15\_1 &      2.4000 &   121 $\pm$     5 &     0.00 $\pm$     0.05 &     5.4 &     8.8 \\
                GOODS\_LRb\_001\_q1\_8\_1 &      2.4284 &  1750 $\pm$    13 &     0.20 $\pm$     0.14 &    19.6 &    10.5 \\
                GOODS\_LRb\_002\_1\_q4\_69\_1 &      2.4304 &    33 $\pm$    32 &     0.25 $\pm$     0.05 &    13.1 &     8.6 \\
                GOODS\_LRb\_dec06\_3\_q3\_53\_3 &      2.4645 &    39 $\pm$     5 &     0.10 $\pm$     0.05 &     4.5 &     8.2 \\
                GOODS\_LRb\_dec06\_1\_q3\_60\_1 &      2.4834 &    13 $\pm$     6 &     0.15 $\pm$     0.05 &    10.9 &     8.1 \\
                GOODS\_LRb\_002\_1\_q4\_1\_1 &      2.4835 &   121 $\pm$     7 &     0.05 $\pm$     0.05 &    22.1 &     9.4 \\
                GOODS\_LRb\_dec06\_3\_q3\_60\_1 &      2.5144 &     9 $\pm$    18 &     0.40 $\pm$     0.05 &    11.9 &     8.0 \\
                GOODS\_LRb\_002\_1\_q1\_26\_1 &      2.5600 &     1 $\pm$     5 &     0.40 $\pm$     0.16 &    13.4 &     7.1 \\
                GOODS\_LRb\_001\_q3\_47\_2 &      2.5641 &   383 $\pm$     9 &     0.00 $\pm$     0.05 &     4.8 &     9.2 \\
                GOODS\_LRb\_001\_q2\_14\_1 &      2.5671 &   215 $\pm$     9 &     0.15 $\pm$     0.05 &     9.2 &     9.2 \\
                GOODS\_LRb\_002\_1\_q4\_32\_1 &      2.6159 &   305 $\pm$     8 &     0.00 $\pm$     0.05 &     7.6 &     9.3 \\
                GOODS\_LRb\_002\_1\_q1\_62\_1 &      2.6191 &   341 $\pm$     8 &     0.00 $\pm$     0.05 &     7.8 &     9.4 \\
                GOODS\_LRb\_001\_q2\_37\_1 &      2.6489 &    77 $\pm$    10 &     0.10 $\pm$     0.05 &    12.5 &     8.9 \\
                GOODS\_LRb\_dec06\_1\_q3\_39\_2 &      2.6607 &     5 $\pm$     6 &     0.25 $\pm$     0.07 &    11.4 &     7.7 \\
                GOODS\_LRb\_dec06\_1\_q3\_32\_1 &      2.6692 &   153 $\pm$    13 &     0.10 $\pm$     0.05 &     8.0 &     9.0 \\
                GOODS\_LRb\_001\_q3\_69\_1 &      2.6796 &     7 $\pm$     4 &     0.10 $\pm$     0.05 &     4.6 &     7.5 \\
                GOODS\_LRb\_001\_q2\_31\_1 &      2.6914 &  1210 $\pm$    10 &     0.10 $\pm$     0.05 &     9.1 &    10.0 \\
                GOODS\_LRb\_001\_q3\_60\_1 &      2.6933 &   171 $\pm$    11 &     0.20 $\pm$     0.05 &     7.2 &     9.0 \\
                GOODS\_LRb\_dec06\_2\_q3\_44\_1 &      2.7034 &   305 $\pm$    14 &     0.15 $\pm$     0.05 &    16.0 &     9.6 \\
                GOODS\_LRb\_001\_q2\_39\_1 &      2.7184 &   101 $\pm$     6 &     0.05 $\pm$     0.05 &     6.4 &     8.8 \\
                GOODS\_LRb\_001\_q2\_44\_2 &      2.7237 &     3 $\pm$     2 &     0.30 $\pm$     0.19 &   104.6 &     8.4 \\
                GOODS\_LRb\_dec06\_1\_q4\_5\_2 &      2.7269 &     3 $\pm$     5 &     0.30 $\pm$     0.08 &    10.7 &     7.5 \\
                GOODS\_LRb\_001\_q3\_9\_1 &      2.8079 &    27 $\pm$    10 &     0.30 $\pm$     0.05 &    10.7 &     8.4 \\
                GOODS\_LRb\_001\_q2\_9\_1 &      2.8121 &    61 $\pm$     9 &     0.15 $\pm$     0.05 &    25.6 &     9.1 \\
                GOODS\_LRb\_dec06\_2\_q3\_51\_1 &      2.8130 &   101 $\pm$    10 &     0.10 $\pm$     0.05 &     9.1 &     8.9 \\
                GOODS\_LRb\_001\_q4\_22\_1 &      2.9659 &   429 $\pm$     3 &     0.00 $\pm$     0.05 &     3.2 &     9.1 \\
                GOODS\_LRb\_001\_q2\_13\_1 &      3.0058 &     5 $\pm$     2 &     0.25 $\pm$     0.19 &   119.2 &     8.7 \\
                GOODS\_LRb\_001\_q3\_48\_1 &      3.0201 &     9 $\pm$     6 &     0.30 $\pm$     0.10 &    16.4 &     8.1 \\
                GOODS\_LRb\_001\_q3\_89\_2 &      3.0300 &   121 $\pm$    11 &     0.15 $\pm$     0.05 &    10.6 &     9.1 \\
                GOODS\_LRb\_dec06\_1\_q4\_7\_1 &      3.0872 &    21 $\pm$    27 &     0.20 $\pm$     0.07 &    12.3 &     8.4 \\
                GOODS\_LRb\_002\_1\_q1\_52\_2 &      3.0920 &    39 $\pm$     8 &     0.10 $\pm$     0.05 &     3.9 &     8.1 \\
                GOODS\_LRb\_dec06\_1\_q4\_45\_3 &      3.1316 &   121 $\pm$     9 &     0.10 $\pm$     0.05 &     6.7 &     8.9 \\
                GOODS\_LRb\_001\_q2\_84\_1 &      3.1677 &     9 $\pm$     6 &     0.15 $\pm$     0.05 &    12.1 &     8.0 \\
                GOODS\_LRb\_001\_q3\_84\_1 &      3.1733 &   763 $\pm$    10 &     0.05 $\pm$     0.05 &     6.4 &     9.6 \\
                GOODS\_LRb\_dec06\_3\_q3\_42\_1 &      3.2000 &   193 $\pm$    18 &     0.10 $\pm$     0.05 &     6.8 &     9.1 \\
                GOODS\_LRb\_dec06\_1\_q3\_41\_1 &      3.2078 &    69 $\pm$    10 &     0.10 $\pm$     0.05 &    27.0 &     9.2 \\
                GOODS\_LRb\_dec06\_3\_q3\_65\_2 &      3.2293 &    57 $\pm$    16 &     0.20 $\pm$     0.05 &    21.5 &     9.0 \\
                GOODS\_LRb\_001\_q3\_71\_2 &      3.2479 &    87 $\pm$    25 &     0.15 $\pm$     0.06 &     9.3 &     8.9 \\
                GOODS\_LRb\_002\_1\_q1\_33\_1 &      3.3565 &     5 $\pm$     4 &     0.20 $\pm$     0.05 &     8.7 &     7.6 \\
                GOODS\_LRb\_001\_q2\_30\_1 &      3.3850 &   153 $\pm$     6 &     0.05 $\pm$     0.05 &     8.1 &     9.0 \\
                GOODS\_LRb\_002\_1\_q2\_71\_1 &      3.4724 &    25 $\pm$     9 &     0.15 $\pm$     0.05 &    21.5 &     8.7 \\
                GOODS\_LRb\_dec06\_3\_q3\_31\_1 &      3.5400 &    37 $\pm$     6 &     0.10 $\pm$     0.05 &     4.6 &     8.2 \\

\hline                                   %inserts single line
\end{tabular}\label{tabla_LAES_properties}
\end{table*}

\end{document}